# Simulation of acquisition shifts in T2 Flair MR images to stress test AI segmentation networks

Christiane Posselt,[a] Mehmet Yigit Avci,[b] Mehmet Yigitsoy,[b] Patrick Schünke,[c] Christoph Kolbitsch,[c] Tobias Schäffter,[c,d] Stefanie Remmele[a*]

[a]University of Applied Sciences, Faculty of Electrical and Industrial Engineering, Am Lurzenhof 1, Landshut, Germany

[b]deepc GmbH, Blumenstrasse 28, 80331 Munich, Germany

[c]Physikalisch Technische Bundesanstalt, Abbestrasse 2-12, 10587 Berlin, Germany

[d]Technical University of Berlin, Department of Medical Engineering, Dovestrasse 6, Berlin, Germany

*corresponding author: Stefanie Remmele, E-mail: Stefanie.Remmele@haw-landshut.de

**Abstract**

**Purpose:** To provide a simulation framework for routine neuroimaging test data, which allows for "stress testing" of deep segmentation networks against acquisition shifts that commonly occur in clinical practice for T2 weighted (T2w) fluid attenuated inversion recovery (FLAIR) Magnetic Resonance Imaging (MRI) protocols.

**Approach:** The approach simulates "acquisition shift derivatives" of MR images based on MR signal equations. Experiments comprise the validation of the simulated images by real MR scans and example stress tests on state-of-the-art MS lesion segmentation networks to explore a generic model function to describe the F1 score in dependence of the contrast-affecting sequence parameters echo time (TE) and inversion time (TI)

**Results:** The differences between real and simulated images range up to 19% in gray and white matter for extreme parameter settings. For the segmentation networks under test the F1 score dependency on TE and TI can be well described by quadratic model functions ($R^2 > 0.9$). The coefficients of the model functions indicate that changes of TE have more influence on the model performance than TI.

**Conclusions:** We show that these deviations are in the range of values as may be caused by erroneous or individual differences of relaxation times as described by literature. The coefficients of the F1 model function allow for quantitative comparison of the influences of TE and TI. Limitations arise mainly from tissues with the low baseline signal (like CSF) and when the protocol contains contrast-affecting measures that cannot be modelled due to missing information in the DICOM header.

# 1 Introduction

In recent years, several machine learning and deep learning (DL) technologies have passed the approval process for a medical device to support radiologists in the diagnosis of medical images[1]. Still, the reliability of these new medical software devices and the underlying DL networks strongly depends on the training data and how well they represent the variety of real clinical image data (test images). Castro et al., describe different sources of "shifts" between training and test domains and among those, the "acquisition shift, resulting from the use of different scanners or imaging protocols, which is one of the most notorious and well-studied sources of dataset shift in medical imaging"[2]. Accordingly, several institutions underline the need for test procedures and published concepts for the evaluation of the robustness and transferability of a model to other data domains[3–6]. The ECLAIR guidelines[6], for example, request „to check robustness to variability of acquisition parameters"[6]. This is especially important for Magnetic Resonance Imaging (MRI), because MR acquisition protocols typically have a large number of sequence parameters, which affect the contrast, resolution and SNR of the acquired images. On the one hand, this allows a wide range of clinical information to be presented by MR images, but on the other hand, it leads to a high heterogeneity between different radiology centers. MR acquisition protocols are often optimized individually at each site and sometimes even for different patients to take patient specific features (e.g. weight and size) into account. Thus, acquisition parameters may vary even for the same type of scan, hence resulting in different image contrasts. There are guidelines providing recommendations on appropriate MR protocols. Among those, e.g. the recently published MAGNIMS–CMSC–NAIMS consensus guidelines[7] prescribe the contrast weighting (i.e. T2w, T2w FLAIR, contrast enhanced T1w) of the scans that need to be included in the "Recommended Core" of protocols for the examination of patients with multiple sclerosis (MS). Nevertheless, they lack specific information on contrast-affecting parameters such as echo, repetition and inversion time (TE, TR, TI) etc.

A multitude of visualization methods have been developed to identify the features within images that a neural network is most sensitive to[8]. Other methods quantify the uncertainty of a network during image processing[9]. However, there is no test procedure that predicts whether an artificial intelligence (AI) product can be applied to the images of a particular radiology practice without loss of performance, e.g. given their customized imaging protocols. Further, it is currently not possible to determine, which acquisition parameters can be changed without compromising the performance of an AI product.

The identification of the influencing factors that a system is most prone to is a well-known problem in the field of process improvement and quality management. It is generally solved by systematic testing based on the "Design of Experiment (DoE)" concept. DoE is a standardized statistical tool for quality control in Six Sigma processes to systematically evaluate the robustness of a process to its influencing factors (see Ref. 10, Chapter 5.4). It predicts the minimum number of experiments needed to quantify and compare the impact of all influencing factors and their interactions on a system's outcome or performance metric. Combined with dedicated analysis of the results, the dominating factors can be easily identified. However, to optimize the experimental design to the given problem, regression analysis needs to be performed to estimate the underlying model function that quantifies the dependence of the response variable (here: AI network performance) on the process' input (here: acquisition parameters), see Ref. 10, Chapter 5.3.3.6.

Therefore, the foremost objective of this work is to study the dependency of a network to the most relevant contrast-affecting acquisition parameters. In the above-mentioned neuroimaging T2w FLAIR scans for example, the echo time (TE) and the inversion time (TI) have the strongest influence on the imaging contrast. But how can models be validated against the typical MR protocol variability of routine scans or even stress tested against rare but realistic maximum domain shifts if the related data are not available?

The benchmark dataset CLEVR-XAI aims to create a selective, controlled and realistic test environment for the evaluation of explainable neural networks in non-medical applications[11]. Similar projects for medical applications have just started[12]. Using machine learning and neural networks for the simulation and synthesis of medical images is a field of intense research. E.g. attempts have already been made to recreate MRI images through simulation and synthesis, e.g. using generative adversarial networks (GANs) or variational autoencoders (VAEs), phantoms and dedicated multi-parametric MR sequences[13]. Other simulators use virtual phantoms, for example from Brainweb and Shepp-Logan, which represent the human brain[14,15] to generate images that represent a particular protocol. The limiting factors in all the above mentioned approaches, however, are either the limited number of anatomies (Brainweb) the lack of anatomical realism (shepp-logan), the dependency on specific software (sequences) or hardware (phantoms) or the ability to synthesize the result of arbitrary MRI sequences settings with only one model (GANs etc.).

The secondary objective of this study is thus the target-oriented combination of simulation and synthesis to generate artificial MRI data of arbitrary sequence character (i.e. "shift derivatives") from a set of real MR images. These data are finally used to stress test a model against variations of acquisition parameters.

For the sake of simplicity, the experiments in this study are focusing on the simulation of domain shift derivatives of T2w FLAIR scans for different TE and TI values to describe the performance of MS lesion segmentation networks in dependence of these scan parameters.

## 2 Methods

This work comprises two levels of methodology and experiments (see Table 1). First, the simulation of domain shift derivatives given a real baseline data set and second, the use of these data to stress test state-of-the-art (SOTA) MS lesion segmentation networks against these shifts. Those networks are trained on data (Table 2) of heterogeneous contrast (e.g. from different field strengths and using different acquisition protocols). The stress tests then intend to show to what extend the networks are robust to changes of image contrast. The simulated data are validated by real MRI scans. The dependency of the models' performance (F1-score) against changes of the MRI protocol parameters (TI, TE) is modelled by second order polynomial functions, recommended by the above-mentioned DoE-guidelines to quantitatively compare the robustness of the networks against acquisition shifts, by the functions' coefficients.

**Table 1** Research questions - methodology - experiments

| Research Questions | General Methodology | Experiments |
|---|---|---|
| | | |

| How well can acquisition shift derivatives of a (real) MRI data set be modelled? | 1. Estimation of tissue properties (tissue segmentation, partial volume tissue fractions $PV$, relaxation parameters $\vec{p}_{Relax}$, texture map $S_{Tex}$). <br> 2. Simulation of acquisition shifts based on MRI signal equation dependent on arbitrary sequence parameters $\vec{p}_{Seq} = (TI, TE)$. | 1:1 Comparison of simulated and real MRI scans in healthy volunteers by average MR signal values in gray matter, white matter and CSF. Comparison of the heuristic estimates of T1 and T2 in tissue ROIs with those of relaxometry methods and literature. |
|---|---|---|
| Is a quadratic model function appropriate to describe the dependence of the F1 score of a segmentation network to acquisition shifts (i.e. changing sequence parameters)? | 1. Generation of representative shift derivatives of a real MS data set. <br> 2. Measurement of F1 as a function of $\vec{p}_{Seq} = (TI, TE,)$ in model tests. | Modelling of $F1(\vec{p}_{Seq})$ as a second order polynomial function, using $R^2$ as a metric to evaluate the model fit. |

The MS data used in this study consist of several open MRI benchmark data sets (see Table 2).

**Table 2.** Data sets used in this work. The simulations utilize the first data set (OpenMS* longitudinal) as baseline data since all contrast-affecting parameters (TE, TI, TR) are provided for these data.

|  | Data | Nr | Description | Source |
|---|---|---|---|---|
|  |  |  |  |  |

| | Dataset | N | Imaging parameters | Scanner |
|---|---|---|---|---|
| Baseline and test data | OpenMS* (longitudinal)[16] | 20 | 2D FLAIR Image: TR = 11000 ms, TE = 140 ms, TI = 2800 ms, FA = 90°, sampling: 0.9 × 0.9 × 3 mm³ | 1.5 T Philips, University Medical Centre Ljubljana (UMCL) |
| Training data | OpenMS (crosssectional)[17] | 30 | 3D FLAIR image: TR = 5000 ms, TE = 392 ms, TI = 1800 ms, FA = 120°, sampling: 0.47 × 0.47 × 0.80 mm³ | 3 T Siemens Magnetom Trio, University Medical Center Ljubljana (UMCL) |
| | Lesion Challenge 2015[18] | 5 | 2D FLAIR Image: TI = 835 ms, TE = 68 ms, sampling: 0.82 × 0.82 × 2.2 mm³ | 3 T Philips, Best, The Netherlands |
| | Lesion Segmentation Challenge 2008[19] | 20 | 2D FLAIR Image: sampling: 0.5 × 0.5 × 0.5 mm³ | 3 T Siemens |
| | MSSEG-2[20] | 40 | 3D FLAIR image | 1.5 T & 3 T GE, Philips, Siemens |
| | NAMIC[21] | 4 | 3D FLAIR image: sampling: 1 × 1 × 1 mm³ | 3 T Siemens Magnetom Trio |

*2.1 The concept of image generation to mimic acquisition shifts*

Data simulation uses an in-vivo MRI scan (baseline data) and mimics changes of that baseline scan in response to changing sequence parameters. The concept of image generation is based on Eq. 1.

$$S(\vec{r}) = \kappa \cdot \left( \left( \sum_{t=1}^{Nr\ Tissues} PV_t(\vec{r}) \cdot s_{FLAIR,t}(\vec{p}_{Tis,t}, \vec{p}_{Seq}) \right) + S_{Tex}(\vec{r}) \right)$$

$$= \left( \left( \sum_{t=1}^{Nr\ Tissues} PV_t(\vec{r}) \cdot \kappa \cdot s_{FLAIR,t}(\vec{p}_{Tis,t}, \vec{p}_{Seq}) \right) + \kappa \cdot S_{Tex}(\vec{r}) \right),$$

(1)

with $S(\vec{r})$, being the simulated signal at pixel position $\vec{r} = (x, y, z)$. The contribution $s_{FLAIR,t}$ of each tissue t to the signal of a pixel or voxel is weighted with its local volume fraction $PV_t(\vec{r})$. $\kappa$ is the (typically unknown) DICOM scaling factor. The texture map $S_{Tex}(\vec{r})$ is introduced to approximate all texture influences other than tissue, e.g. based on artifacts, field inhomogeneities, noise, etc. The entire image generation process therefore consists of two different steps (Fig. 1). The first step comprises the preliminary estimation of these tissue properties followed by the second step, the final image simulation according to Eq. 1.

$s_{FLAIR,t}(\vec{p}_{Tis,t}, \vec{p}_{Seq})$ is the signal as determined by the sequence and the tissue properties, i.e. the parameters $\vec{p}_{Tis,t} = (\rho_t, T1_t, T2_t)$ of the underlying tissue t (like the spin density $\rho$ and relaxation parameters T1 and T2 of gray matter (GM), white matter (WM), cerebrospinal fluid (CSF) and lesion). $s_{FLAIR,t}$ is given by the T2w FLAIR signal equation given in Eq. 2 as published in Ref. 22.

$$s_{FLAIR,t}(\vec{p}_{Tis,t}, \vec{p}_{Seq}) = \rho_t \cdot \left(1 - 2 \cdot \exp\left(-\frac{TI}{T1_t}\right) + \exp\left(-\frac{(TR-TE_{last})}{T1_t}\right)\right) \cdot \exp\left(-\frac{TE}{T2_t}\right) \quad (2)$$

with $\vec{p}_{Seq} = (TE, TI, TR, ...)$, i.e. the sequence parameters.

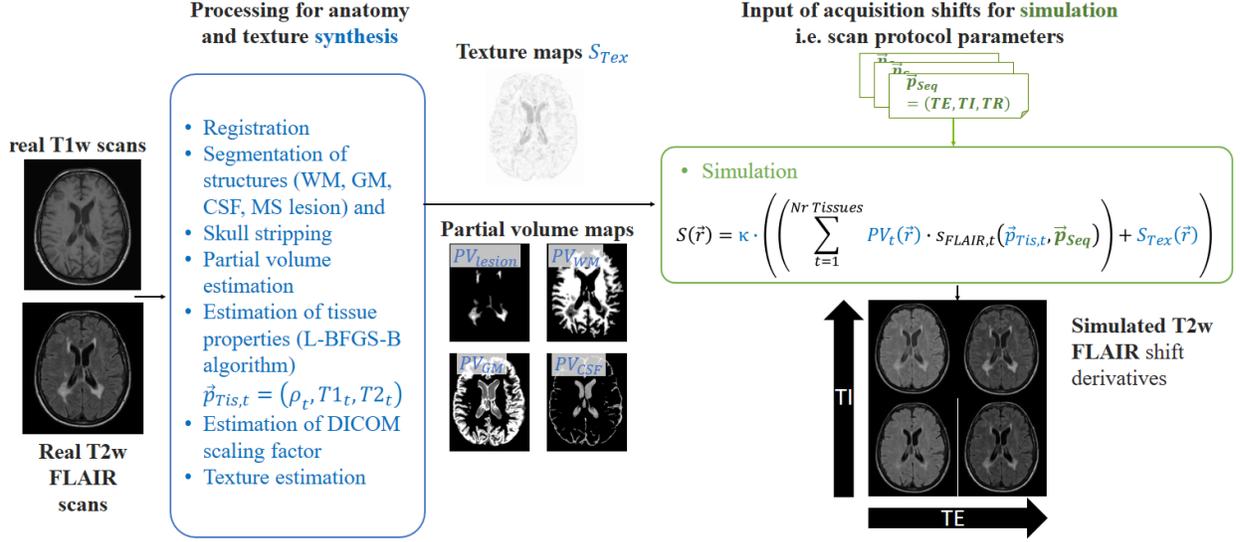

**Fig. 1** Acquisition shifts of a real baseline data set are simulated based on the MRI signal equation of a T2w FLAIR sequence. The signal contribution of each tissue t is scaled by its volume fraction $PV_t$ and enriched by a texture map $S_{tex}$. All influences other that of the sequence (anatomic structures, DICOM scaling or texture) are synthesized (blue box) from the real baseline scan prior to simulation (green box).

*2.1.1 Simulation and synthesis methods*

The equations 1 and 2 contain a number of tissue parameters that must be represented as realistic as possible for the data generation process but cannot be easily simulated (e.g. anatomical structures, lesion sizes and locations). The idea behind the proposed generative approach is thus to combine image synthesis and simulation as follows:

1. Synthesis: anatomy and disease related signal contributions are derived from a real MR baseline data set $S_m$. These data are used to estimate:

   — the partial volume maps $PV_t(\vec{r})$ by using a partial volume estimation method based on Ref. 23 (see next section). The approach requires an additional T1w scan, which is also included in the above-mentioned "Recommended core" protocols of MS examinations. The approach further requires prior tissue segmentation.

— the DICOM scaling factor $\kappa$ of the baseline T2w FLAIR scan and

— $S_{Tex}(\vec{r})$, to mimic other texture influences (e.g. noise, artifacts).

2. Simulation of all signal contributions that are affected by the sequence and the choice of parameters.

— Simulation of acquisition shifts is performed through variation of $\vec{p}_{Seq}$ in $s_{FLAIR,t}(\vec{p}_{Tis,t}, \vec{p}_{Seq})$ using Eq. 2. $T1$ and $T2$ are set to random values within a realistic range.

*2.1.2 Partial volume estimation*

For estimation of the partial volume fractions of each tissue we apply the method described in Ref. 23. This approach requires that a signal rise or decline from one region to the other is unique for one kind of tissue-tissue interface. However, when the brain contains lesions, a rise of signal when leaving the WM region may be attributed to either a WM-lesion OR an WM-GM interface. The partial volume maps are thus generated in two steps, assuming that lesions are solely located in and surrounded by WM[24]. First, as required by the approach, segmentation masks are created. We were using Synthseg[25] for segmentation of normal tissues, expert lesion masks were provided through the data sets[26]. Second, the T1w scans are used to estimate the PV-maps $PV_{WM1}$, $PV_{GM}$ and $PV_{CSF}$ of normal tissue. Lesion pixels might be falsely assigned to the PV-map of GM, which can be easily corrected for by setting the GM maps to 0 at all lesion pixels as given by segmentation. Third, the FLAIR images, masked to contain only WM and lesions, are fed through the PV-algorithm, to obtain another $PV_{WM2}$ and $PV_{lesion}$ map. The final $PV_{WM}$ is initialized with $PV_{WM1}$. Then, in those pixels, where $PV_{lesion} > 0$, the partial volume fraction in WM is set to $PV_{WM} = 1 - PV_{lesion}$. All steps are summarized in Fig. 2.

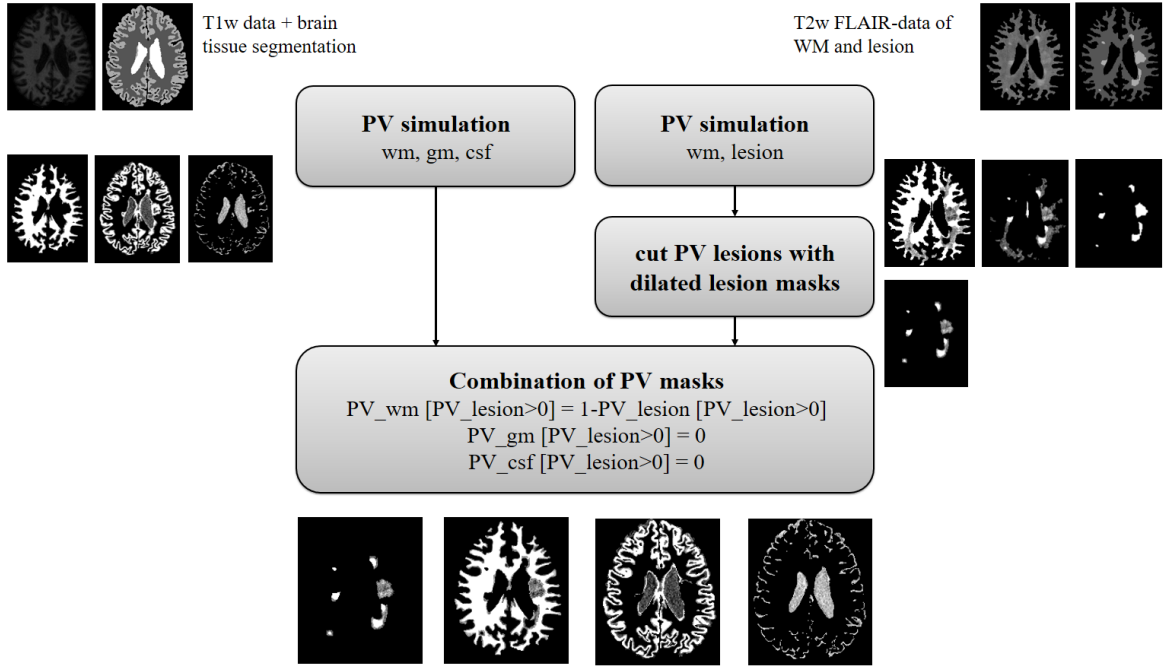

**Fig. 2** Partial volume maps (PV) for normal tissue are determined based on a T1w scan and the method described in Ref. 23. The PV map for the lesion is estimated using the same method and a WM-lesion segment of the T2w FLAIR scan, where lesions differentiate better from the WM background. Fusing all PV information yields the final PV maps.

*2.1.3 Estimation of the DICOM scaling factor κ and the texture map $S_{Tex}(\vec{r})$*

A simplified version of Eq. 1 describes the signal of those pixels of the real baseline image $S_m$ that contain only one tissue fraction ($PV = 1$)

$$S_{m,t}(\vec{r}) = \underbrace{\kappa \cdot 1 \cdot s_{FLAIR,t}(\vec{p}_{Tis,t}, \vec{p}_{Seq})}_{signal\ term} + \underbrace{\kappa \cdot S_{Tex}(\vec{r})}_{texture\ term} \text{ for all } \vec{r}_{PVt=1}, \text{ where } PV_t(\vec{r}) = 1. \quad (3)$$

Since both $S_{Tex}$ and $\vec{p}_{Tis,t}$ are unknown, the problem to compute $S_{Tex}$ is overdetermined. We solve this by introducing the assumption that signal variations are primarily caused by noise and thus the average texture $\bar{S}_{Tex}(\vec{r}_{PVt=1})$ in this region is 0. Eq. 3 can then be written as

$$\bar{S}_{m,t}(\vec{r}_{PVt=1}) = \kappa \cdot s_{FLAIR,t}(\vec{p}_{Tis,t}, \vec{p}_{Seq}). \quad (4)$$

This allows for a preliminary estimation of the apparent tissue parameters $\tilde{\vec{p}}_{Tis,t}$ from the ratio of average real and simulated signals for different tissues t (the ratio eliminates the unknown $\kappa$ in Eq. 4), or more precisely by comparing the real and simulated contrast metrics given in Eq. 5 and 6.

$$C_{sim,t_1,t_2} = \frac{s_{FLAIR,t_1}(\vec{p}_{TisEst,t_1},\vec{p}_{Seq}) - s_{FLAIR,t_2}(\vec{p}_{TisEst,t_2},\vec{p}_{Seq})}{s_{FLAIR,t_1}(\vec{p}_{TisEst,t_1},\vec{p}_{Seq}) + s_{FLAIR,t_2}(\vec{p}_{TisEst,t_2},\vec{p}_{Seq})} \quad (5)$$

$$C_{m,t_1,t_2} = \frac{\bar{S}_{m,t_1}(\vec{r}_{PVt_1=1}) - \bar{S}_{m,t_2}(\vec{r}_{PVt_2=1})}{\bar{S}_{m,t_1}(\vec{r}_{PVt_1=1}) + \bar{S}_{m,t_2}(\vec{r}_{PVt_2=1})}. \quad (6)$$

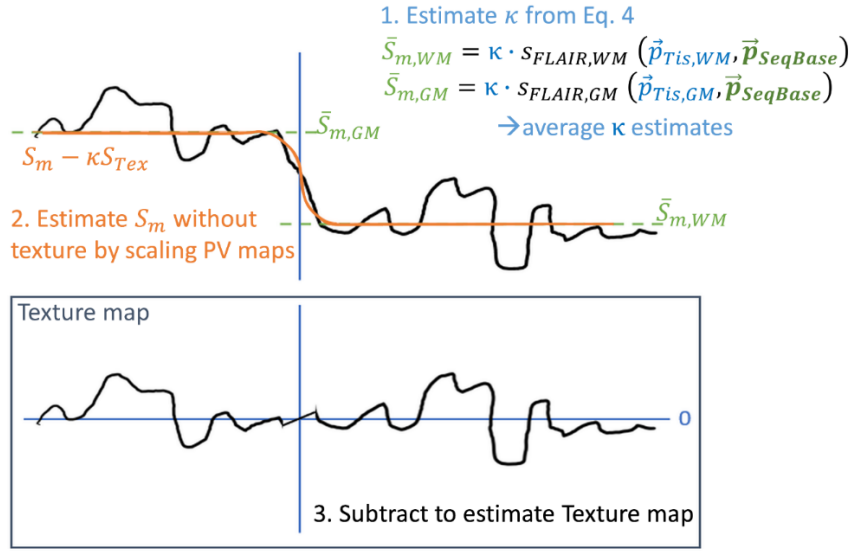

**Fig. 3** Method to estimate the texture of an MR image by subtracting the estimated signal in consideration of the partial volume effect.

The parameters of $\vec{p}_{Tis,t}$ are optimized to minimize the cost function

$$\left(C_{sim,GM,WM} - C_{m,GM,WM}\right)^2 + \left(C_{sim,CSF,WM} - C_{m,CSF,WM}\right)^2 + \left(C_{sim,Lesion,WM} - C_{m,Lesion,WM}\right)^2 \rightarrow min \quad (7).$$

Then, $\kappa$ can be estimated using Eq. 4. Now, that all unknowns are determined, Eq. 1 is solved to determine the texture map $S_{Tex}(\vec{r})$ (see Fig. 3).

*2.1.4 Experiments - Comparison of simulation and measurement.*

MR images of 10 healthy volunteers were acquired to compare the simulations with real measurements. The examinations were approved by the ethics committee of the Physikalisch-Technische Bundesanstalt (PTB) and are in accordance with the relevant guidelines and regulations. Written informed consent was obtained from all volunteers prior to the measurements. Data were acquired at 3T (Siemens Verio) using the following sequences: a magnetization prepared rapid gradient echo (MPRAGE) for the estimation of the PV-maps (3D, TR = 2300 ms, TI = 900 ms, TE = 3.2 ms, voxel size: 0.75 × 0.75 × 4.69 mm³) and five T2w FLAIR scans as a reference measurement for the simulated images (Multislice 2D, TR = 9000 ms, voxel size: 0.75 × 0.75 × 4.69 mm³) with TE and TI values as given in Table 3. to represent the extreme shift derivatives of the possible scan domain and its center (see Fig. 5). The "center" protocol serves as the baseline scan for the simulations of the "corner" protocols.

**Table 3** TE and TI of the five T2w FLAIR acquisition protocols

| TE / ms | TI / ms |
|---------|---------|
| 112     | 2500    |
| 84      | 2200    |
| 84      | 2900    |
| 150     | 2200    |

| 150 | 2900 |

Reference T1 values were obtained from saturation-recovery measurements. Eleven T1-weighted images for different saturation delay times (TD = 0.1, 0.2, 0.3, 0.4, 0.5, 0.75, 1.0 1.25, 1.5, 2.0, 8.0 s) were acquired using a fully sampled single-shot centric-reordered GRE readout (TE/TR = 3.0/6.5 ms, flip angle = 6°, voxel size: 1.3 x 1.3 x 8.0 mm³) implemented in pulseq[27]. Final quantitative T1 values were generated using a non-linear least squares curve fitting algorithm[28] assuming a simple mono-exponential magnetization recovery. T2 reference values were derived from the two different echo times ($TE_1$ = 84 ms and $TE_2$ = 150 ms) of the FLAIR scans $S_m$ using Eq. 8

$$T_2 = \frac{TE_2 - TE_1}{\ln\left(\frac{S_m(TE_1)}{S_m(TE_2)}\right)}. \quad (8)$$

The T2 estimates obtained with TI = 2900 ms and 2200 ms are averaged to deliver the final reference T2 values. The relaxometry estimates described in section 2.1.3 are compared to these reference values and to values given by literature[29–32]. Finally, the five real and simulated scans are compared by the theoretical percentage signal deviation per ms relaxometry errors $\Delta T1$ and $\Delta T2$ approximated by error propagation as

$$ds_{T1}(T1) = \frac{\frac{ds_{FLAIR}(T1,T2)}{dT1}}{s_{FLAIR}} \cdot 100\%, ds_{T2}(T2) = \frac{\frac{ds_{FLAIR}(T1,T2)}{dT2}}{s_{FLAIR}} \cdot 100\% \quad (9)$$

and in dependence of T1 and T2 to confirm that signal differences are related to relaxometry imperfections. The stress test pipeline is summarized in Fig. 4 and comprises two steps as described in the following sections.

*2.2 Model stress tests to determine the influence of acquisition shifts*

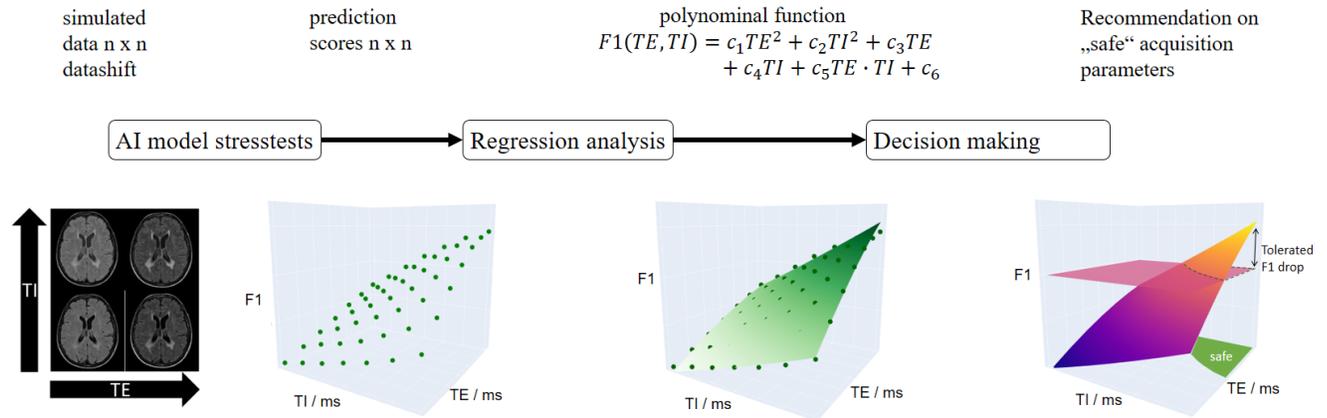

**Fig. 4** The AI model undergoes testing using generated images that represent varying acquisition shifts. Regression analysis delivers a model function for F1 to provide the user with an assessment of the model's limitations.

*2.2.1 Generation of test data*

With the methods described in 2.1, derivatives of the baseline data can be generated that represent arbitrary acquisition shifts of a baseline scan ("shift derivatives"). Commonly used variations of scan protocols (minimum and maximum TE and TI values) were estimated using literature and real scans. The outcome of that investigation is published in Ref. [33] and is depicted in Fig. 5. 7x7 test data sets were generated that represent 7 different TE values and 7 different TI values, since these are the most contrast-affecting parameters in T2w FLAIR sequences.

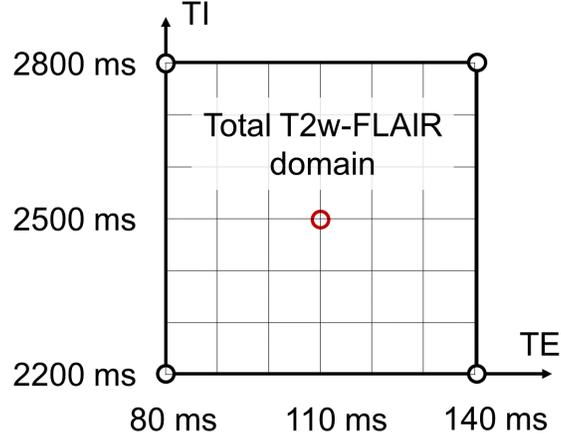

**Fig. 5** Minimum and maximum values of TI and TE as determined by literature research and real scans. These values limit the real-world scan domain. Test data are generated by simulation to represent all possible data within this domain on a regular grid. The corners and the center (red circle) determine the MRI protocols for reference measurements used to validate the simulated data.

*2.2.2 Modelling the network performance in dependence of sequence parameters*

The lesion F1 score of a lesion segmentation network can be determined for all of these data comparing the network prediction with the lesion ground truth segmentation masks. Averaging all lesion F1 scores finally delivers F1 as a function of TE and TI. We use a response surface method (quadratic model, cubic terms neglected) to describe the dependence of F1 on arbitrary values of the influencing factors TE and TI and their interactions as recommended by Ref. 10. Accordingly, the quadratic model

$$F1(TE, TI) = c1 \cdot TE^2 + c2 \cdot TI^2 + c3 \cdot (TI \cdot TE)^2 + c4 \cdot TE + c5 \cdot TI + c6 \cdot TI \cdot TE + c7$$

(10)

is fitted to these F1 measurements. The coefficients $c1$ to $c6$ can be understood as measure for the relevance of the influencing factors $TE$ and $TI$ (main factors) and their interactions $TE \cdot TI$.

*2.2.3 Experiments - stress testing SOTA models against acquisition shifts.*

To validate the model function described in Eq. 10 two state-of-the-art (SOTA) models are trained on data with heterogeneous contrast as described in Table 2. First, the nnU-Net framework is used, which utilizes an U-Net architecture and automatically configures its hyperparameters and configuration[34]. The first model is a 3D full resolution nnU-Net, which is chosen by nnU-Net's auto-configured framework as the best performing model among 2D and low-resolution 3D counterparts. Training is done by nnU-Net's self-configured automatic framework. The second model is a SegResNet model, which uses ResNet-like blocks and skip connections without the variational autoencoder part[35]. The network is trained with 64x64x64 cropped blocks for 1000 epochs with an Adam optimizer and learning rate of 0.001 with Pytorch and MONAI sofwares.

The "longitudinal" OpenMS data set is the only open benchmark data set for which all contrast-affecting parameters (TE, TI, TR) are provided (Table 2). All data are skull stripped using the FSL Brain Extraction Tool (FSL BET)[36] prior to all processing steps. The average F1 is determined and modelled as a function of TE and TI as described in 2.2.2. $R^2$ is used to evaluate the appropriateness of the model function in Eq. 10.

# 3   Results

*3.1 Comparison of simulation and measurement*

Fig. 6 shows the variation of the estimated and reference relaxation measurements in comparison to the literature ranges. The estimated and measured relaxation times mostly lie within the literature range. As further underlined by the mean relaxometry values in Table 4, the long T1 value and the low FLAIR signal hampers relaxometry in CSF. The literature does not report on CSF T2 measurements at 3T. T2 is independent of the field strength but even at 1.5 T, to our knowledge, the Brainweb catalogue is the only literature source reporting a T2 value for CSF (329 ms), although the values presented in that catalogue (in WM and GM) tend to be lower than most other values at 1.5T[37].

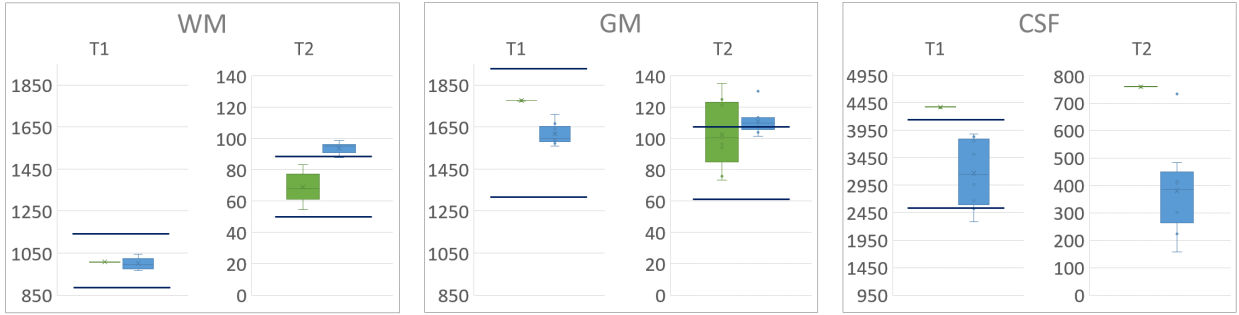

**Fig. 6** Comparison of the ranges of estimated (green) and reference relaxation measurements (blue) at 3T. Blue lines show the literature ranges[29–32].

Table 4. Mean values for T1 and T2 in normal tissue. All values are given in ms

|  | $T1_{wm}$ | $T2_{wm}$ | $T1_{gm}$ | $T2_{gm}$ | $T1_{csf}$ | $T2_{csf}$ |
|---|---|---|---|---|---|---|
| **estimated** | 1007 ± 1 | 69 ± 9 | 1776 ± 3 | 102 ± 20 | 4376 ± 4 | 760 ± 379 |
| **measured** | 999 ± 27 | 94 ± 3 | 1616 ± 46 | 111 ± 8 | 3176 ± 568 | 379 ± 157 |

Visually, simulation and measurement agree well (Fig. 7). Small scaling errors of the nulled CSF signal result in high relative signal deviations. In addition, Table 5 lists the relative error between real and simulated images in different manually drawn ROIs.

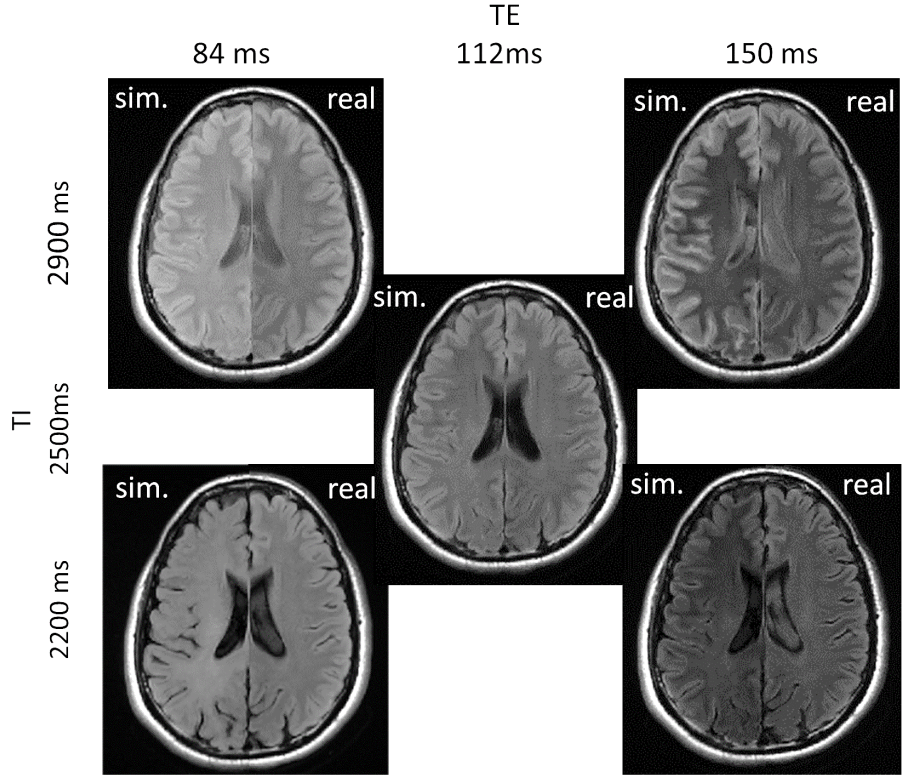

**Fig. 7** MRI simulations (left side of the MRIs) and their real counterparts (right side of the MRIs) for all five protocols of one example volunteer. The simulation results are embedded in the skull segment to adjust the scaling of the images.

**Table 5** Comparison of the mean signals of WM, GM, CSF and skull of simulation and reference MRI with relative error in %.

| TE (ms) | TI (ms) | WM | GM | CSF |
|---|---|---|---|---|
| 150 | 2900 | 18% ± 6% | 7% ± 3% | 75% ± 30% |
| 150 | 2200 | 19% ± 6% | 9% ± 7% | 36% ± 9% |
| 112 | 2500 | 0% ± 0% | 0% ± 0% | 0% ± 0% |
| 84 | 2900 | 13% ± 6% | 8% ± 6% | 22% ± 13% |
| 84 | 2200 | 12% ± 6% | 8% ± 5% | 58% ± 10% |

The deviation between simulation and measured MR signals is higher in WM as in GM. The theoretical error propagation of the relaxometry estimates on the simulated signal is depicted in Fig. 8.

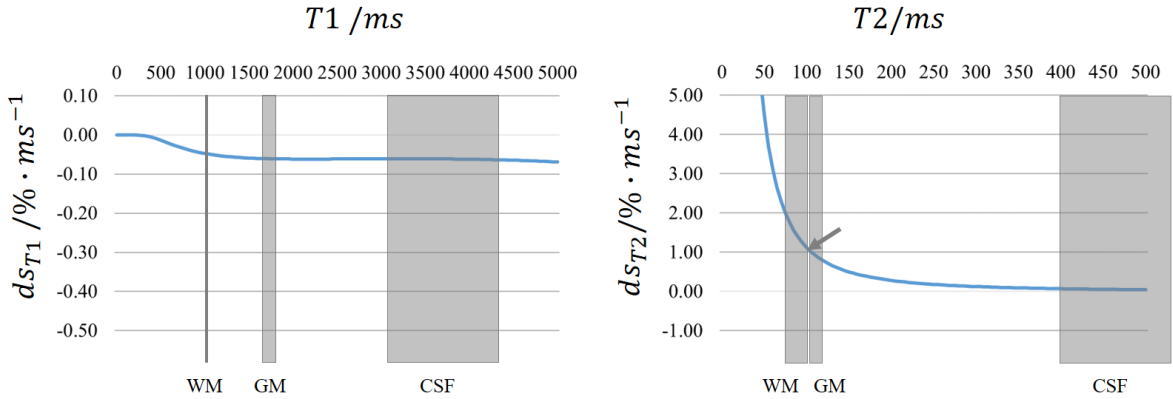

**Fig. 8** Percentage signal simulation errors per ms relaxometry value as described by error propagation in dependence on the tissue's T1 or T2 values. E.g. (see arrows) the overestimation of T2 by 1ms results in about 1% signal simulation error of WM and GM signals (here: given average protocol parameters). The absolute errors increase with T1 and decrease with T2.

*3.2 Results of stress testing SOTA models against acquisition shift*

Testing the models with the real baseline data and their simulated counterpart (TE = 140 ms and TI = 2800 ms) yields F1 scores, which differ in the fourth decimal place (OpenMS data: SegResNet: 0.4398 ± 0.2242; nnU-Net: 0.6105 ± 0.1500, see Fig. 9). The coefficient of determination $R^2$ of the model fit (2nd order polynomial) is 0.991 for the SegResNet results and 0.982 for the nnU-Net results.

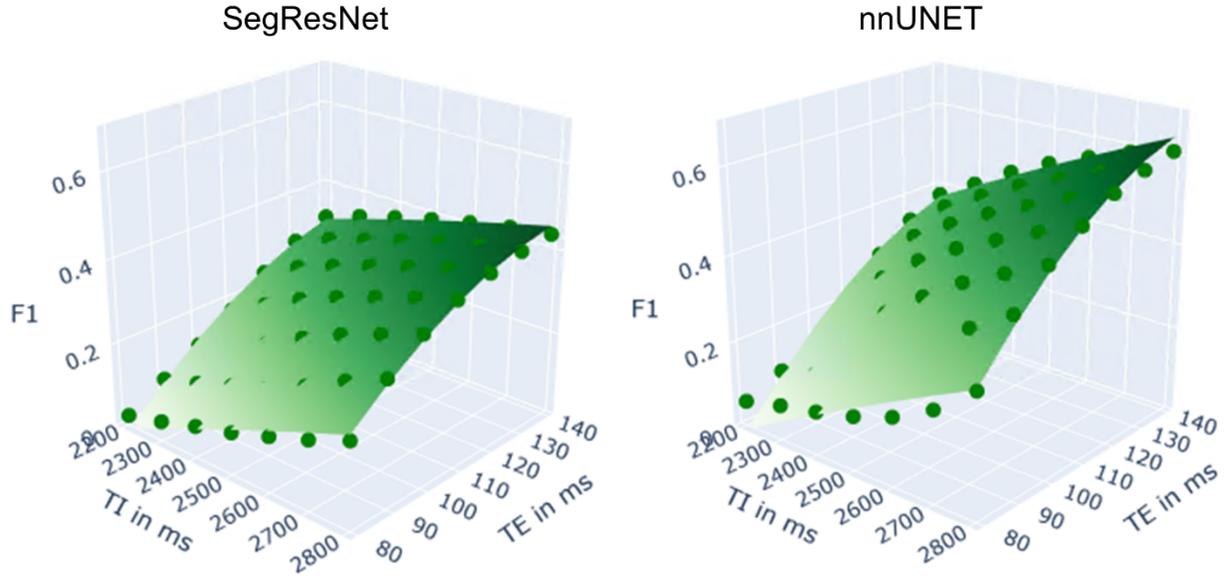

**Fig. 9** The surface plots show the behavior of the AI models in dependence of the data shifts. Points: F1 scores of the predictions, surface: model fit, i.e. the F1 trend as a function of the acquisition parameters TE and TI.

The coefficients for Eq. 10 in Table 6 show that TE has the highest influence on both segmentation network.

**Table 6:** Coefficients of model fit c1 to c7 of Eq. 10. Units are given in $ms^{-1}$ and $ms^{-2}$ for linear, quadratic and combined terms, respectively. The highest coefficients are those, scaling the influencing factor TE.

|  | intersection | TE | TI | TE² | TI² | TE·TI |
| --- | --- | --- | --- | --- | --- | --- |
| **nnU-Net** | $-2.95$ | $\mathbf{2.24 \cdot 10^{-2}}$ | $9.42 \cdot 10^{-4}$ | $-6.48 \cdot 10^{-5}$ | $-7.59 \cdot 10^{-8}$ | $-8.14 \cdot 10^{-7}$ |
| **SegResNet** | $-2.57$ | $\mathbf{2.14 \cdot 10^{-2}}$ | $8.25 \cdot 10^{-4}$ | $-5.15 \cdot 10^{-5}$ | $-5.37 \cdot 10^{-8}$ | $-2.18 \cdot 10^{-6}$ |

In the simulated images of Fig. 10, the lesion-to-WM contrast decreases for lower TE and TI values. This is accompanied by a decline of the F1-score, i.e. the models' ability to differentiate between lesion and white matter decreases with lower contrast.

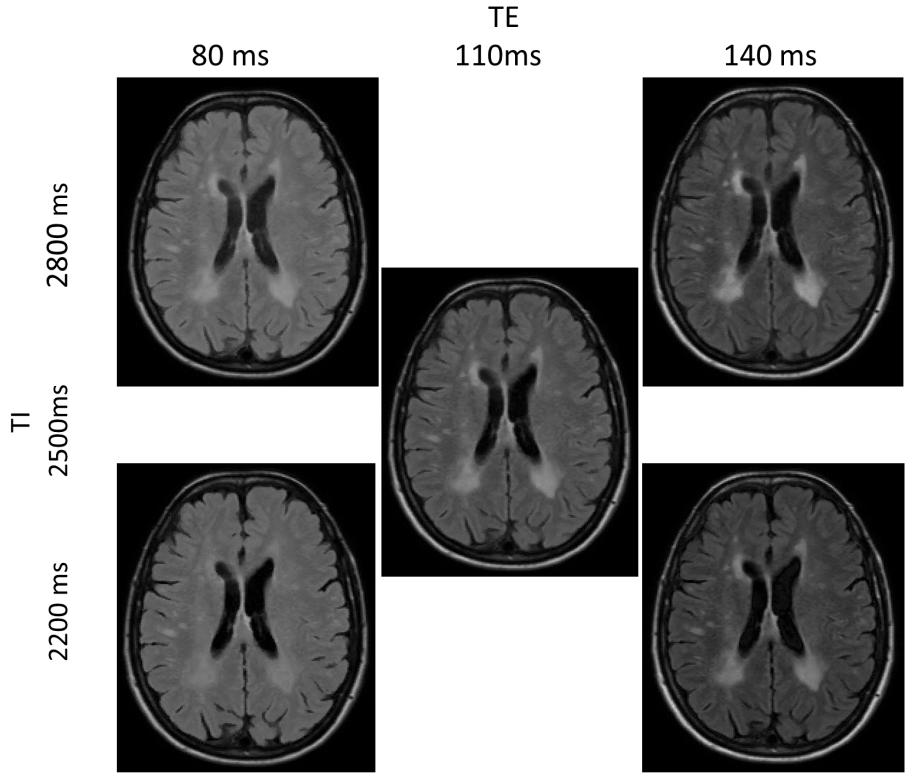

**Fig. 10** MRI simulations for five protocols (Fig. 5) of one patient of the OpenMS dataset. The simulation results are embedded in the skull segment to adjust the scaling of the images.

## 4 Discussion and Conclusion

The image generation method simulates acquisition shift derivatives of a real baseline scan for arbitrary sequence parameters. It was designed to be applicable to common clinical neuroimaging studies that normally contain T2w FLAIR and T1w images. It does not require extra sequences but only knowledge on the scan parameters of the baseline T2w FLAIR data.

*4.1 Comparison of simulation and measurements*

At the extreme points of the experimental design the simulation shows a 19% deviation to the measured values in white matter and lower deviation in gray matter. This can most likely be explained by the inaccuracies of the relaxometry method used in this work. Using the error propagation as a rough guess, the misestimation of 19% could be explained by a 19 ms deviation of T2, which is likely to be realistic considering the reference measurements and the range of literature reference values. And even those reference relaxometry methods suffer from inaccuracies caused by inflow or sequence imperfections, in particular when estimating the T1 and T2 of flowing tissue like blood or CSF[38]. One could improve the validation by including T1 and T2 mapping sequences in the same resolution and spatial coverage. Common relaxometry sequences in neuroimaging rely on multiple 3D spoiled gradient recalled echo or inversion recovery sequences for T1 mapping and multi-echo or balanced steady-state free precession sequences at variable flip angles for T2 mapping[38,39]. The imaging study in this work was already time consuming due to the 5-times repetition of the lengthy T2w FLAIR protocol and the T1 weighted scan. Therefore, there was just limited time for a rough dual echo T2 estimation and for addition of a time-efficient single-slice T1-mapping protocol (acquisition time ~ 30 seconds) to examine the T1 estimates in one slice, and thus values were compared ROI-wise. Still, the T1 and T2 values estimated here mostly lie in the range of literature values and differences to the reference measurements are also comparable to the range of literature values. A one-to-one comparison of real and simulated images is challenging as it requires the exact knowledge of the relaxation times of that particular patient. Precise relaxometry is neither the aim of this work nor is it necessary for the simulation of test data. The relaxometry parameters in Eq. 1 and 2 are set to arbitrary values to deliver a representative cohort of anatomies. Relaxometry imperfections hamper accurate validation of the simulated values, yet, they manifest only in a misestimation of the DICOM scaling factor $\kappa$ and thus in under- or overestimation of the texture amplitude. Unfortunately, for MRI sequences this scaling factor is not part of the DICOM header as it is for the Hounsfield Units in CT imaging. Irregularities of the texture amplitude on the other hand, might be balanced by normalizing the texture amplitude over the entire data set. Furthermore, the texture amplitude could be also included as another influencing factor in the stress test analysis in addition to the sequence parameters - e.g. as a measure of noise or artifact level. In contrast to using other AI-based generative approaches like GANs, VAEs or diffusion models[13,40–43] the underlying signal equation allows for the generation of arbitrary but distinct shift derivatives from just one data set.

*4.2 Stress test results*

The F1(TE, TI) measurements seems to be well described by the quadratic function. The metric varies only smoothly so that cubic terms can be neglected. TE seems to be the most influencing factor for all models, which is in line with the nature of the contrast weighting of the sequence (T2w FLAIR).

Furthermore, the lesion F1 values are comparable to that of real data (72 % [44]) at least in or close to the baseline representation. The performance decreases towards the extreme points of the experimental grid. One has to bear in mind that these extreme points are mathematical constraints, given by the minimum and maximum combinations of TE and TI of real sequences. The boundary of the experimental grid does not represent the boundary of the typical scan domain. The latter does not necessarily contain the combination of extreme values of both TE and TI at the same time. Those extreme data simulations are thus not part of the training data therefore causing severe drops in the F1 value.

The high F1 scores for the two „high-TE corners" (Fig. 9) can further be explained by the high lesion contrast for these protocols. In contrast, the low lesion contrast for the low TE, low TI protocol comes with low F1 scores respectively. Another contribution of this work is thus also a proof-of-concept for the description of the performance metric of an AI model in dependence of its influencing factors. The modelling yields a quantitative comparison of the relevance of all influencing factors. This concept of surface response modelling is based on well established experimental designs and could be easily transferred to other common metrics[45] (e.g. confusion matrix and derivatives or even uncertainty estimates[46]) or other models (e.g. classification models). Now, that the model function was confirmed, the number of experiments could by reduced significantly in future studied to reduce the computational effort. For the optimal "positioning" of these sample points on the "domain grid" for meaningful sampling of the surface response curve, state-of-the-art guidelines in the field of experimental design offer several recommendations depending on the number of influencing factors[10].

*4.3 Limitations*

One important limitation is the small number of test data sets used in this study. Thus, the absolute results of the stress tests might not be representative for a larger cohort of patients and lesions. They can only serve as a sample domain grid to confirm an appropriate model function and to demonstrate the proof-of-concept. Unfortunately, all open MS data are provided in nifti format and the OpenMS data are the only data that come at least with the information on TE, TI and TR and thus all sequence parameters needed in the simulation. In real world applications, one can assume that manufacturers of models have access to the entire DICOM header that also includes tags for TE, TI, TR and many more. Thus, in theory, more acquisition shifts caused by other sequence parameters could be incorporated as influencing factors in the stress tests. However, since the number of sampling points on the domain grid quickly rises with every new influencing factor, a prior prioritization is crucial.

An intrinsic limitation of the T2w FLAIR and T1w sequences is that the CSF signal is very low or even nulled hampering partial volume estimation and relaxometry in this tissue. Accordingly, the differences between the simulations and measurements become most apparent in CSF compared to the other tissues, limiting the validation of the approach in CSF. Future work should investigate, if tissue and relaxometry estimation can be improved by additionally incorporating the contrast of conventional T2w sequences in the first step of the image generation pipeline as in these images, CSF shows up brightly. All three scans (T2w, T2w FLAIR and the post Gd T1w scan) constitute the "Recommended Core" in current MS scanning guidelines[7].

Another limitation is the assumption that the average texture contribution to the signal is zero. This is not true in the case of artifacts resulting from inhomogeneities of B0, B1 or the receive coil sensitivity profile[47,48]. The method is further only applicable to baseline images, of which the contrast can be fully described by the parameters accessible in the DICOM header. E.g. the parameter $TE_{last}$ in Eq. 2 is approximated by $2 \cdot TE$, since it is not part of the DICOM header. In the real measurements in volunteers, the true value for $TE_{last}$ was 30 % higher. In these experiments, changing the parameter to the correct value did not have any influence on the outcome of the comparison (due to the long TR value). Still, there might be other measures of contrast manipulation in T2w FLAIR studies that are not accessible by the DICOM tags and that prevent an accurate estimation of the DICOM scaling factor and thus the texture amplitude (e.g. modulated RF pulses to prevent the signal from decaying in long echo trains, acceleration techniques and dedicated k-space ordering, particularly common in 3D sequences[22,49–52], blood inflow[53], etc). Future work should elaborate to what extend these influences and their extents can be modelled and incorporated either in the simulation e.g. by random guesses or in the stress tests represented by additional influencing factors.

Despite these limitations, the presented framework, containing image simulation and stress test methodology allows for investigation of the robustness of AI models in response to arbitrary data shifts. Accordingly, due to the lack of a gold standard, the metrological proof of the F1 response to parameter changes is not possible and absolute predictions about these values remain uncertain. However, influencing parameters in the MR sequence can be compared with each other by the surface model coefficients and - given a tolerated performance drop - "safe" parameters settings can be at least roughly assessed (Fig. 4). Using the simulation algorithm as an alternative augmentation method also allows for introducing a-priori knowledge on MR signal variations into the AI-model development process.

*Disclosures*

*Code, Data, and Materials Availability*

The MS data utilized in this study are listed Table 2. The data policy of the clinical study does not allow free access to the volunteer MRI data. Due to the collaboration agreement with the industrial partner the code cannot be made available.

*Acknowledgments*

This project was funded by the Zentrales Innovationsprogramm Mittelstand (ZIM, KK5050201 LB0). This work has further been partly supported by Collaborative Research Centre "Matrix in Vision" funded by German Research Foundation (DFG CRC-1340).

**Stefanie Remmele** is a professor for medical technologies at the University of Applied Sciences in Landshut, Germany since 2012. Prior to that, she conducted research on quantitative MR methods at Philips Research in Hamburg, Germany. She received her Diploma in Electrical Engineering and her PhD on MR methods from the Karlsruhe Institute of Technology (KIT) in 2003 and in 2006, respectively. Her current research interests include image simulation and synthesis in radiology and image guided therapy. She is a member of SPIE.

**Caption List**

Table 1 Research questions - methodology - experiments

Table 2. Data sets used in this work. The simulations utilize the first data set (OpenMS* longitudinal) as baseline data since all contrast-affecting parameters (TE, TI, TR) are provided for these data.

Table 3 TE and TI of the five T2w FLAIR acquisition protocols

Table 4. Mean values for T1 and T2 in normal tissue. All values are given in ms

Table 5 Comparison of the mean signals of WM, GM, CSF and skull of simulation and reference MRI with relative error in %.

Table 6: Coefficients of model fit c1 to c7 of Eq. 10. Units are given in $ms-1\ and\ ms-2$ for linear, quadratic and combined terms, respectively. The highest coefficients are those, scaling the influencing factor TE.

Fig. 1 Acquisition shifts of a real baseline data set are simulated based on the MRI signal equation of a T2w FLAIR sequence. The signal contribution of each tissue t is scaled by its volume fraction $PVt$ and enriched by a texture map $Stex$. All influences other that of the sequence (anatomic structures, DICOM scaling or texture) are synthesized (blue box) from the real baseline scan prior to simulation (green box).

Fig. 2 Partial volume maps (PV) for normal tissue are determined based on a T1w scan and the method described in Ref. 23. The PV map for the lesion is estimated using the same method and a WM-lesion segment of the T2w FLAIR scan, where lesions differentiate better from the WM background. Fusing all PV information yields the final PV maps.

Fig. 3 Method to estimate the texture of an MR image by subtracting the estimated signal in consideration of the partial volume effect.

Fig. 4 The AI model undergoes testing using generated images that represent varying acquisition shifts. Regression analysis delivers a model function for F1 to provide the user with an assessment of the model's limitations.

Fig. 5 Minimum and maximum values of TI and TE as determined by literature research and real scans. These values limit the real-world scan domain. Test data are generated by simulation to represent all possible data within this domain on a regular grid. The corners and the center (red circle) determine the MRI protocols for reference measurements used to validate the simulated data.

Fig. 6 Comparison of the ranges of estimated (green) and reference relaxation measurements (blue) at 3T. Blue lines show the literature ranges[29–32].

Fig. 7 MRI simulations (left side of the MRIs) and their real counterparts (right side of the MRIs) for all five protocols of one example volunteer. The simulation results are embedded in the skull segment to adjust the scaling of the images.

Fig. 8 Percentage signal simulation errors per ms relaxometry value as described by error propagation in dependence on the tissue's T1 or T2 values. E.g. (see arrows) the overestimation of T2 by 1ms results in about 1% signal simulation error of WM and GM signals (here: given average protocol parameters). The absolute errors increase with T1 and decrease with T2.

Fig. 9 The surface plots show the behavior of the AI models in dependence of the data shifts. Points: F1 scores of the predictions, surface: model fit, i.e. the F1 trend as a function of the acquisition parameters TE and TI.

Fig. 10 MRI simulations for five protocols (Fig. 5) of one patient of the OpenMS dataset. The simulation results are embedded in the skull segment to adjust the scaling of the images.